\documentclass[superscriptaddress,reprint]{revtex4-1}
\usepackage[utf8]{inputenc}
\usepackage{upgreek}
\usepackage{graphicx}
\usepackage[english]{babel}
\usepackage{amsmath}
\usepackage{gensymb} 

\usepackage[hidelinks]{hyperref}
\usepackage[all]{hypcap} 

\begin{document}
\author{Zoltán Kovács-Krausz}
\affiliation{Department of Physics, Budapest University of Technology and Economics and Nanoelectronics 'Momentum' Research Group of the Hungarian Academy of Sciences, Budafoki ut 8, 1111 Budapest, Hungary}
\author{Anamul Md Hoque}
\affiliation{Department of Microtechnology and Nanoscience, Chalmers University of Technology, SE-41296, Göteborg, Sweden}
\author{Péter Makk}
\email{E-mail: peter.makk@mail.bme.hu}
\affiliation{Department of Physics, Budapest University of Technology and Economics and Nanoelectronics 'Momentum' Research Group of the Hungarian Academy of Sciences, Budafoki ut 8, 1111 Budapest, Hungary}
\author{Bálint Szentpéteri}
\affiliation{Department of Physics, Budapest University of Technology and Economics and Nanoelectronics 'Momentum' Research Group of the Hungarian Academy of Sciences, Budafoki ut 8, 1111 Budapest, Hungary}
\author{Mátyás Kocsis}
\affiliation{Department of Physics, Budapest University of Technology and Economics and Nanoelectronics 'Momentum' Research Group of the Hungarian Academy of Sciences, Budafoki ut 8, 1111 Budapest, Hungary}
\author{Bálint Fülöp}
\affiliation{Department of Physics, Budapest University of Technology and Economics and Nanoelectronics 'Momentum' Research Group of the Hungarian Academy of Sciences, Budafoki ut 8, 1111 Budapest, Hungary}
\author{Michael Vasilievich Yakushev}
\affiliation{M.N. Miheev Institute of Metal Physics of UB RAS, 620108, Ekaterinburg, Russia.}
\affiliation{Ural Federal University, Ekaterinburg, 620002, Russia.}
\affiliation{Institute of Solid State Chemistry of UB RAS, Ekaterinburg, 620990, Russia.}
\author{Tatyana Vladimirovna Kuznetsova}
\affiliation{M.N. Miheev Institute of Metal Physics of UB RAS, 620108, Ekaterinburg, Russia.}
\affiliation{Ural Federal University, Ekaterinburg, 620002, Russia.}
\author{Oleg Evgenevich Tereshchenko}
\affiliation{St. Petersburg State University, 198504, St. Petersburg, Russia.}
\affiliation{A.V. Rzhanov Institute of Semiconductor Physics, 630090, Novosibirsk, Russia.}
\affiliation{Novosibirsk State University, 630090, Novosibirsk, Russia.}
\author{Konstantin Aleksandrovich Kokh}
\affiliation{St. Petersburg State University, 198504, St. Petersburg, Russia.}
\affiliation{Novosibirsk State University, 630090, Novosibirsk, Russia.}
\affiliation{V.S. Sobolev Institute of Geology and Mineralogy, 630090, Novosibirsk, Russia.}
\author{István Endre Lukács}
\affiliation{Center for Energy Research, Institute of Technical Physics and Material Science, H-1121 Budapest, Hungary}
\author{Takashi Taniguchi}
\affiliation{National Institute for Material Science, 1-1 Namiki, Tsukuba, 305-0044, Japan}
\author{Kenji Watanabe}
\affiliation{National Institute for Material Science, 1-1 Namiki, Tsukuba, 305-0044, Japan}
\author{Saroj Prasad Dash}
\email{E-mail: saroj.dash@chalmers.se}
\affiliation{Department of Microtechnology and Nanoscience, Chalmers University of Technology, SE-41296, Göteborg, Sweden}
\author{Szabolcs Csonka}
\affiliation{Department of Physics, Budapest University of Technology and Economics and Nanoelectronics 'Momentum' Research Group of the Hungarian Academy of Sciences, Budafoki ut 8, 1111 Budapest, Hungary}
\title{Electrically Controlled Spin Injection from Giant Rashba Spin-Orbit Conductor BiTeBr}

\begin{abstract}

Ferromagnetic materials are the widely used source of spin-polarized electrons in spintronic devices, which are controlled by external magnetic fields or spin-transfer torque methods. However, with increasing demand for smaller and faster spintronic components, utilization of spin-orbit phenomena provides promising alternatives. New materials with unique spin textures are highly desirable since all-electric creation and control of spin polarization is expected, where the strength, as well as an arbitrary orientation of the polarization, can be defined without the use of a magnetic field. In this work, we use a novel spin-orbit crystal BiTeBr for this purpose. Owning to its giant Rashba spin splitting, bulk spin polarization is created at room temperature by an electric current. Integrating BiTeBr crystal into graphene-based spin valve devices, we demonstrate for the first time that it acts as a current-controlled spin injector, opening new avenues for future spintronic applications in integrated circuits. 

\end{abstract}

\maketitle
\section{Introduction}

Spin-orbit interaction (SOI), the coupling between the spin and the motion of electrons inside an electrostatic potential, is a central concept in contemporary quantum- and spin-based nanoelectronic devices\,\cite{Manchon_2015}. Materials with strong SOI are key building blocks in topological states of matter, such as quantum spin Hall states\,\cite{Kane_QSHE_2005,Bernevig_QSHE_2006,Konig_QSHE_2008}, Majorana bound states\,\cite{Fu_majo_2008,Sau_majo_2010,Mourik_majo_2012,Lutchyn_majo_2018} or spin textures\,\cite{Soumyanarayanan_2016}. The SOI also leads to the emergence of strong spin-valley coupling in transition metal dichalcogenides (TMDs)\,\cite{Schaibley-valley-review}, facilitates control over spin qubits\,\cite{van_der_Wiel_qdot_2002,Kloeffel_qdot_2013}, or can be used to switch the magnetization of a ferromagnetic nanostructure by spin-orbit induced torque (SOT)\,\cite{Chernyshov_sot_2009,Manchon_sot_2019}. The latter can be realized by the creation of current-induced spin polarization in high SOI materials and heterostructures due to the spin Hall effect in bulk materials\,\cite{she-th-sinova,gr-tmdc-she-wte2,gr-tmdc-she-mos2}, Rashba-Edelstein effect at interfaces\,\cite{ree-exp-remr,sanchez-ree-snag-2016,Lesne_lao-sto_2016,Rodriguez_Vega-ree-2017,ree-exp-ws2-bart,benitez-ree-ws2-2019}, and spin-momentum locking phenomenon in topological materials\,\cite{Vaklinova_grti_2016}.

The recently discovered class of semiconductor materials, bismuth tellurohalides (BiTeX, where X is a halogen element) feature a giant Rashba spin splitting of the bulk bands\,\cite{bitei-rashba,Bahramy_bitei_2011,bitebr-intro-th-rashba} as experimentally verified by spin- and angle-resolved photoemission spectroscopy\,\cite{bitei-intro-exp-arpes,bitebr-intro-exp-synth,bitebr-arpes,bitebr-intro-exp-opt,bitebr-effmass-n3d-tanner,bitebr-effmass-n3d-ideue}. This unique spin texture makes them highly desirable for various spintronic applications. Further interesting properties of these highly polar semiconductor materials include the bulk rectification effects\,\cite{bitebr-effmass-n3d-ideue}, pressure-induced topological phase\,\cite{bitei-intro-topo-bahramy,bitei-intro-topo-chen,bitebr-intro-topo-crassee,bitex-topo-th,bitex-intro-topo-sasagawa}, superconductivity\,\cite{bitei-intro-supra-a,bitei-intro-supra-b}, and out-of-plane spin textures caused by coupling to orbital degree of freedom\,\cite{bawden-bitei-spinorbit-texture-2015}.

\begin{figure}[!ht]
    \includegraphics[width=1.0\columnwidth]{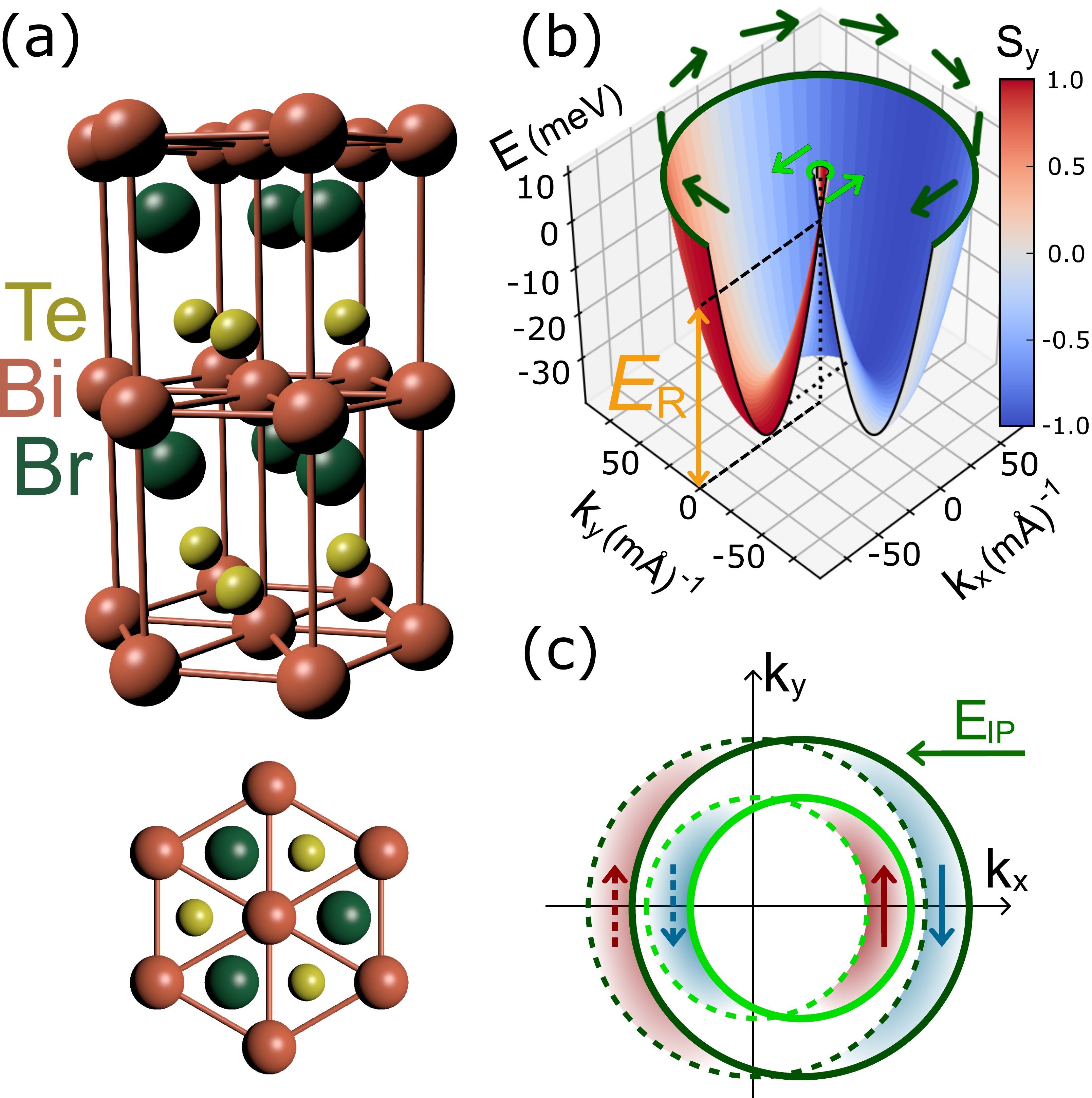}
    \caption{\label{fig0} \textbf{Giant Rashba SOI crystal BiTeBr.} (a) Layered crystal structure of BiTeBr, and top-down view (bottom panel). (b) Calculated Rashba-split conduction band of BiTeBr, with band cut off at the Fermi energy, estimated from the high charge carrier density. Green arrows indicate subband helicity, while the color scale is the y-component of the spin, $S_\text{Y}$. The inner subband, in light green, has opposite helicity compared to the outer one (dark green). (c) Rashba-Edelstein effect depicted in a top-down view of the band structure at Fermi energy, not to scale. An in-plane electric field shifts the occupied states from equilibrium (depicted as dashed circles) by $k_\text{d}=-eE_\text{IP}\tau/\hbar$, where $\tau$ is the momentum scattering time. Due to the intrinsic spin splitting, more states of spin $\downarrow$ are added, corresponding to the spin orientation of the outer subband in the direction of the electric field.}
\end{figure}

The crystal structure of BiTeBr consists of three distinct elemental planes\,\cite{bitebr-structure-popovkin} (see \hyperref[fig0]{Fig.~\ref{fig0}} (a)), with the heavy Bi atoms being located in a z$\rightarrow$-z symmetry breaking built-in electric field. This results in a giant Rashba spin splitting $E_\text{R}\approx$\,40 meV of the subbands (\hyperref[fig0]{Fig.~\ref{fig0}} (b)), which feature spin states perpendicular to momentum with a helical spin structure, opposite in the two subbands. While in equilibrium there is no net spin polarization, an in-plane electric field $E_\text{IP}$ shifts the occupation of states in $k$-space and gives rise to a spontaneous spin polarization near the Fermi level (see \hyperref[fig0]{Fig.~\ref{fig0}} (c)). This shift involves more states on the outer subband than on the inner one, leading to an unbalanced spin population, with more spins oriented along the direction given by the blue arrows compared to the red ones. This current-induced spin polarization mechanism is called the Rashba-Edelstein Effect (REE)\,\cite{ree-aronov,ree-edelstein}, where the magnitude and orientation of spin polarization can be controlled by the strength and direction of the electric field. However, electronic generation of spin polarization in giant Rashba SOI materials, and its utilization for spintronics devices has not been demonstrated so far.

In this work, we report for the first time the all-electrical control over spin polarization in giant Rashba SOI material BiTeBr at room temperature. We couple BiTeBr to a graphene spin-valve device, which allows us to use the very well developed toolbox of graphene spintronics\,\cite{han-grreview-2014,gr-tb-al2o3-tombros,gr-review-ieee,gr-tb-al2o3-popinciuc,gr-tb-transp-han,gr-tb-tio2-han,Roche_2014,drogeler-roomtemp-gr2014}, including ferromagnetic contacts used in spin-sensitive non-local measurements. Detailed measurements and analysis shows that spin current is generated in graphene by REE in the bulk BiTeBr, and demonstrates its possible utilization as an all-electric spin injection source at ambient temperature.

\section{Results and Discussion}

\begin{figure*}[!ht]
    \begin{center}
    \includegraphics[width=1.7\columnwidth]{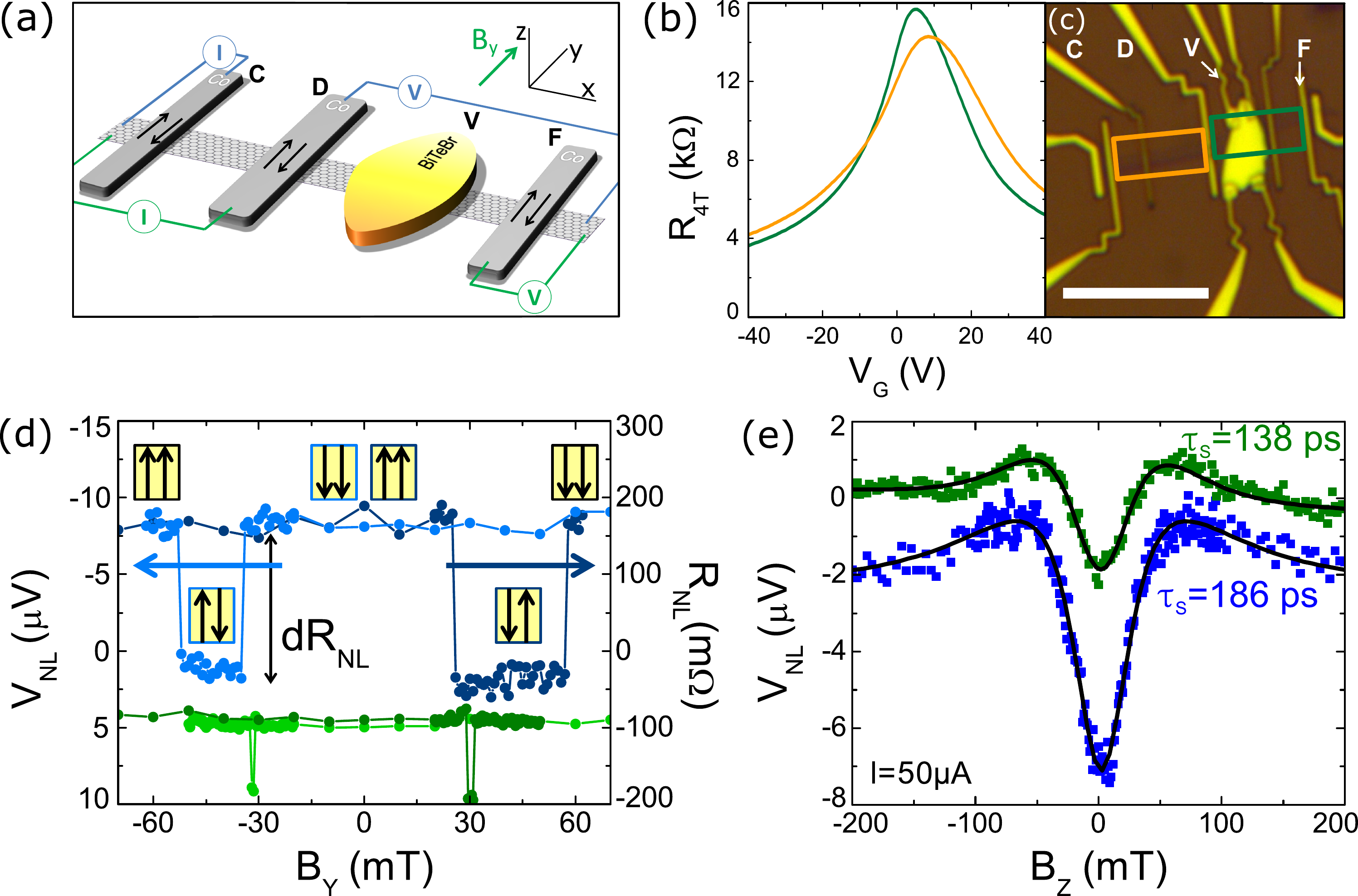}
    \caption{\textbf{Charge and spin transport in the BiTeBr/graphene device.} (a) Schematic of the device, showing the non-local measurement setups for spin transport { in graphene}. { Green is used for the BiTeBr/graphene section, while blue is used as a reference graphene measurement.} (b) Four terminal channel resistance as function of gate voltage for the BiTeBr/graphene and a reference section of identical length { (indicated by the same colors in panel (c))}. Charge transport in the two sections is similar, with a mobility of $\sim$ 2300 and 1800\,cm$^{-2}$(Vs)$^{-1}$, and charge neutrality point $V_{\text{CNP}}$ of 6\,V and 12\,V, for the BiTeBr-containing and reference sections respectively. (c) An optical image of the device, with a 10\,$\upmu$m scale bar. (d) Non-local spin valve measurement of the sections, showing typical switching with magnetic field along FM contact easy axis. Orientations of FM injector-detector pair polarization are shown in the yellow boxes. while horizontal arrows indicate the direction of sweeping magnetic field. (e) Comparison of Hanle spin precession of the BiTeBr/graphene (green) and reference (blue) sections, with extracted spin relaxation times. Data corresponding to the BiTeBr/graphene sections (green) has been scaled up by a factor of 3, and manually offset by +6\,$\upmu$V in (d) and +2\,$\upmu$V in (e) for better visibility. Results indicate the presence of BiTeBr has no significant influence on graphene spin transport properties.}
    \label{fig1}
    \end{center}
\end{figure*}

BiTeBr crystals with a typical thickness of 40-100 nm were integrated in a graphene spin valve consisting of an exfoliated graphene flake and Co ferromagnetic (FM) electrodes (see \hyperref[fig1]{Fig.~\ref{fig1}} (a) and (b)). Graphene serves as an ideal spin transport channel with a spin relaxation length of several microns due to its weak SOI and high mobility\,\cite{Tombros2007}. The magnetization of Co based FM contacts points along the electrode axis (y-direction) due to shape anisotropy. A thin TiO$_2$ tunnel barrier is created at the graphene/FM interface to enhance spin injection and detection efficiency\,\cite{gr-tb-al2o3-tombros,gr-tb-al2o3-popinciuc,gr-tb-tio2-han} (see Methods).

Before turning to REE in BiTeBr, we characterize spin transport behavior and polarization of FM contacts in the graphene spin valve that we use as our spin detector. Spin signal is detected by non-local (NL) spin injection geometry as the blue electric circuit shows on \hyperref[fig1]{Fig.~\ref{fig1}} (a). Current is injected from FM contact C towards the left side of the flake, which induces spin polarization in the graphene. The spin-polarized carriers diffuse towards FM contact D, which depending on the orientation of its magnetization, is sensitive either to spin up or down chemical potential in graphene. Thus measuring a NL voltage, $V_\text{NL}$, between contact D and a distant reference contact,  the spin polarization in graphene can be detected. Blue curves on \hyperref[fig1]{Fig.~\ref{fig1}} (d) show such a NL spin valve measurement as the magnetization orientation of contacts C and D is switched by an external magnetic field, $B_\text{Y}$. Due to the different coercive fields of contacts C and D, their orientation switches from $\uparrow\uparrow$ via $\uparrow\downarrow$ to $\downarrow\downarrow$ as $B_\text{Y}$ is { swept down}. Note that in our device with Co electrodes, the interfacial spin polarization points opposite to the FM magnetization; the black arrows show polarization rather than magnetization. The observed step in $V_\text{NL}$ has a corresponding NL resistance change,  $dR_\text{NL}=dV_\text{NL}/I\approx 190$\,m$\Upomega$ (see \hyperref[fig1]{Fig.~\ref{fig1}} (d)).

Applying an out-of-plane magnetic field, $B_\text{Z}$, the spin relaxation time in the graphene channel can be determined by Hanle spin precession measurements\,\cite{hanle-pathint-jedema}. Here, diffusing spins from C to D also undergo in-plane Larmor precession along with the spin relaxation, resulting in a reduction in $V_\text{NL}$ (see blue curve on \hyperref[fig1]{Fig.~\ref{fig1}} (e)). In addition to the Hanle curve, a small background contribution linear in $B_\text{Z}$, presumably caused by stray charge current, is also visible and included in the fits. Details of the fitting process are described in the Supp. Info.. From fitting (black solid line) the spin relaxation time $\tau_\text{S} = 186$\,ps and spin relaxation length $\lambda_\text{S} = 1.80$\,$\upmu$m were obtained, which are typical values for graphene on SiO$_2$\,\cite{gr-taus-sosenko,gr-taus-kawakami,gr-tb-tio2-han}. The same NL spin valve and Hanle measurements were performed on the graphene channel with BiTeBr crystal on top, as shown by the green circuit on \hyperref[fig1]{Fig.~\ref{fig1}} (a) and green curves on \hyperref[fig1]{Fig.~\ref{fig1}} (d) \& (e). Very similar spin relaxation time $\tau_\text{S} = 138$\,ps and relaxation length $\lambda_\text{S} = 1.64$\,$\upmu$m were determined for the BiTeBr-containing section as for the reference graphene channel.  The smaller $R_\text{NL}$ amplitude for the BiTeBr-containing section is only a consequence of the longer channel length between contacts D-F than between C-D, of 6\,$\upmu$m and 4\,$\upmu$m, respectively. Thus we could conclude that BiTeBr does not significantly alter spin transport in graphene. This is consistent with the similarly insignificant effect of BiTeBr on the charge transport in graphene (see \hyperref[fig1]{Fig.~\ref{fig1}} (b)).

With top contacts fabricated on the BiTeBr crystal (e.g. contact V on \hyperref[fig1]{Fig.~\ref{fig1}} (c)), vertical transport measurements were performed (see Supp. Info. for details), revealing a BiTeBr-graphene interface resistance $R_\text{INT}$ of 10-20\,k$\Upomega$, resistance of the BiTeBr crystal of 100\,Ohm with a very low bulk resistivity on the order of 10$^{-5}$\,$\Upomega$m, and charge carrier density of approximately 10$^{19}$\,cm$^{-3}$, similar to results obtained by Refs.\,\cite{bitebr-effmass-n3d-ideue,bitebr-effmass-n3d-tanner}. The large interface resistance explains the BiTeBr crystal's lack of influence on graphene spin and charge transport. It also makes an ideal configuration for spin injection into graphene, due to avoiding conductance mismatch\,\cite{vanwees-mismatch-2000,fert-mismatch-2001,Tombros2007} between graphene and the highly conductive BiTeBr.
 
\begin{figure*}[!ht]
    \begin{center}
    \includegraphics[width=1.5\columnwidth]{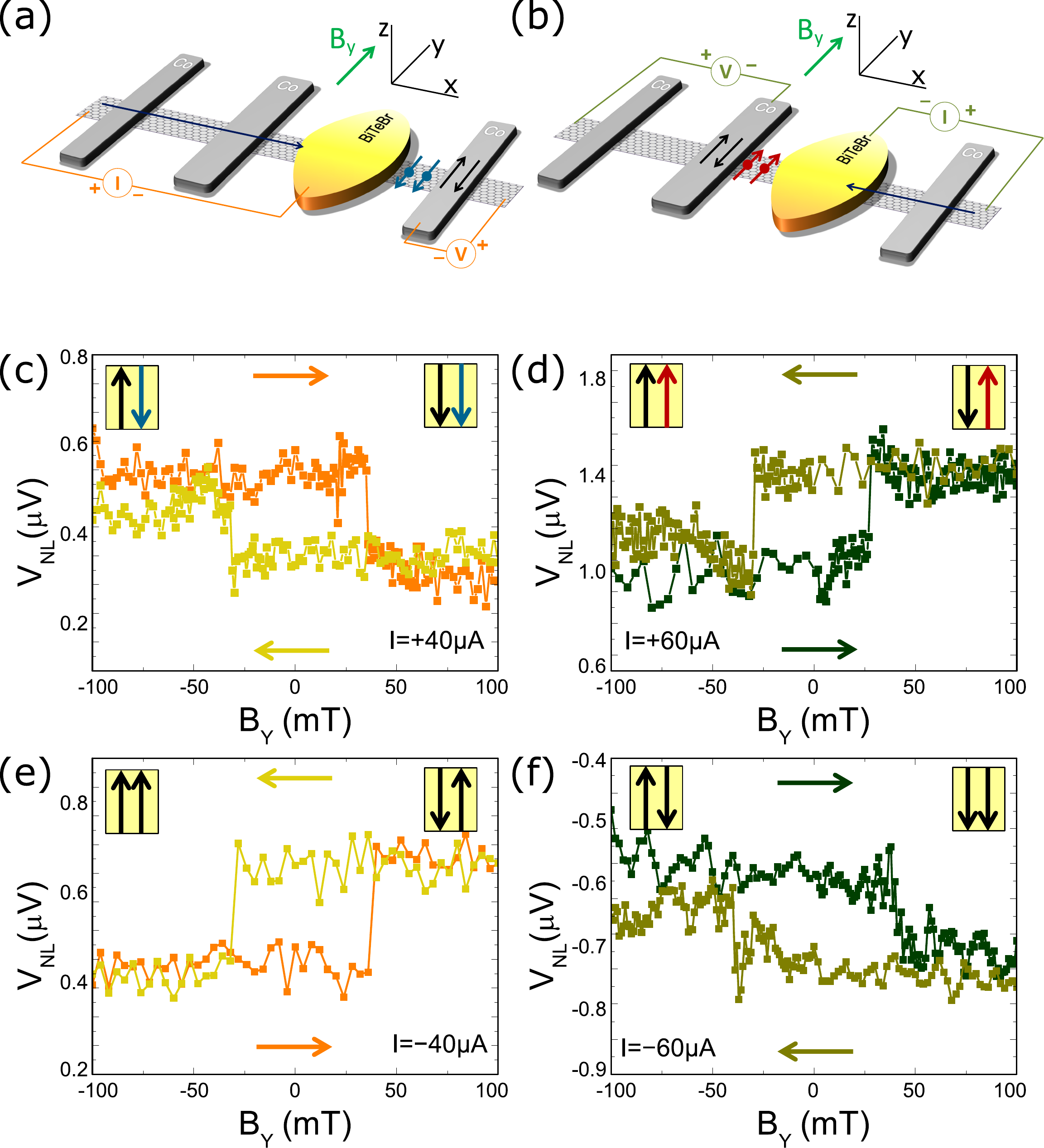}
    \caption{\textbf{Spin injection from BiTeBr to graphene and its dependence on electric field orientation.} (a) \& (b) { Measurement schematics of spin injection from BiTeBr crystal, and non-local detection using FM contact on opposite sides of the device, to demonstrate bias orientation dependence.} (c) \& (d) Spin-polarized signal detected non-locally using the setup in (a) \& (b) respectively, using positive bias current. Horizontal arrows indicate the direction of magnetic field sweep. The injected spin polarization is opposite on (c) compared to (d). (e) \& (f) The same measurement using negative bias current. The observed spin polarization injected into graphene is the same as in (c) \& (d). The parallel and antiparallel spin configurations are indicated in the yellow boxes.}
    \label{fig2}
    \end{center}
\end{figure*}

Now we will use the graphene channel in an unconventional NL spin valve configuration, where BiTeBr crystal serves as an \textit{injector electrode}, using the previously characterized FM contacts as detector. Bias current passing through the crystal can facilitate spontaneous spin polarization through REE in BiTeBr, and the current transports the polarized charge carriers into graphene, where they diffuse toward the detector. BiTeBr-injected spin polarization is observed by using two setups as depicted in \hyperref[fig2]{Fig.~\ref{fig2}} (a) \& (b), where in (a) the FM detector is on the right of the BiTeBr and the current sink is on the left, and vice versa in (b). Because the electric field orientation is different, the spins injected in (a) shown in blue will have a different orientation from those in (b), shown in red. This is fundamentally different from a FM injector contact, where the spin orientation does not depend on electric field orientation in this way. \hyperref[fig2]{Fig.~\ref{fig2}} (c) depicts the spin signal observed by setup (a). In contrast  to standard spin valve measurements with two FM contacts (see \hyperref[fig1]{Fig.~\ref{fig1}} (d)), here we only observe a single switch - instead of two - in the NL voltage as $B_\text{Y}$ is swept up (orange) or down (yellow). The position of this switch ($B_\text{Y}\approx \pm33\,\text{mT}$), corresponds to the switching field of the FM detector, determined in previous spin valve measurements. At negative $B_\text{Y}$, from the increase in $V_\text{NL}$ after this switch, one can conclude that the spin orientation injected from BiTeBr (blue arrows on panel (a) and (c)) becomes antiparallel with FM detector polarization.

The lack of a second switching in $V_\text{NL}$ is consistent with REE effect in bulk BiTeBr, since spin polarization injected in this way will not be affected by the applied magnetic field. At positive $B_\text{Y}$, $V_\text{NL}$ shows a lower value, where detector polarization and direction of injected spins becomes parallel again. The same NL measurement was carried out by using a FM detector and current sink on the opposite side of BiTeBr, setup (b). Compared to setup (a), the NL voltage now decreases as $B_\text{Y}$ is reduced (olive curve), which corresponds to reversed orientation of BiTeBr-injected spins (red arrow). This is also in agreement with REE, where a opposite spin polarization is expected if the electric field is flipped. In addition, the current direction was also reversed for both NL geometries (panels (e) and (f)), which changes the sign of the $V_\text{NL}$ jump in both cases, as is expected from REE. In terms of $dR_\text{NL}$, the detected values are 5-10\,m$\Upomega$.

\begin{figure*}[!ht]
    \begin{center}
    \includegraphics[width=1.6\columnwidth]{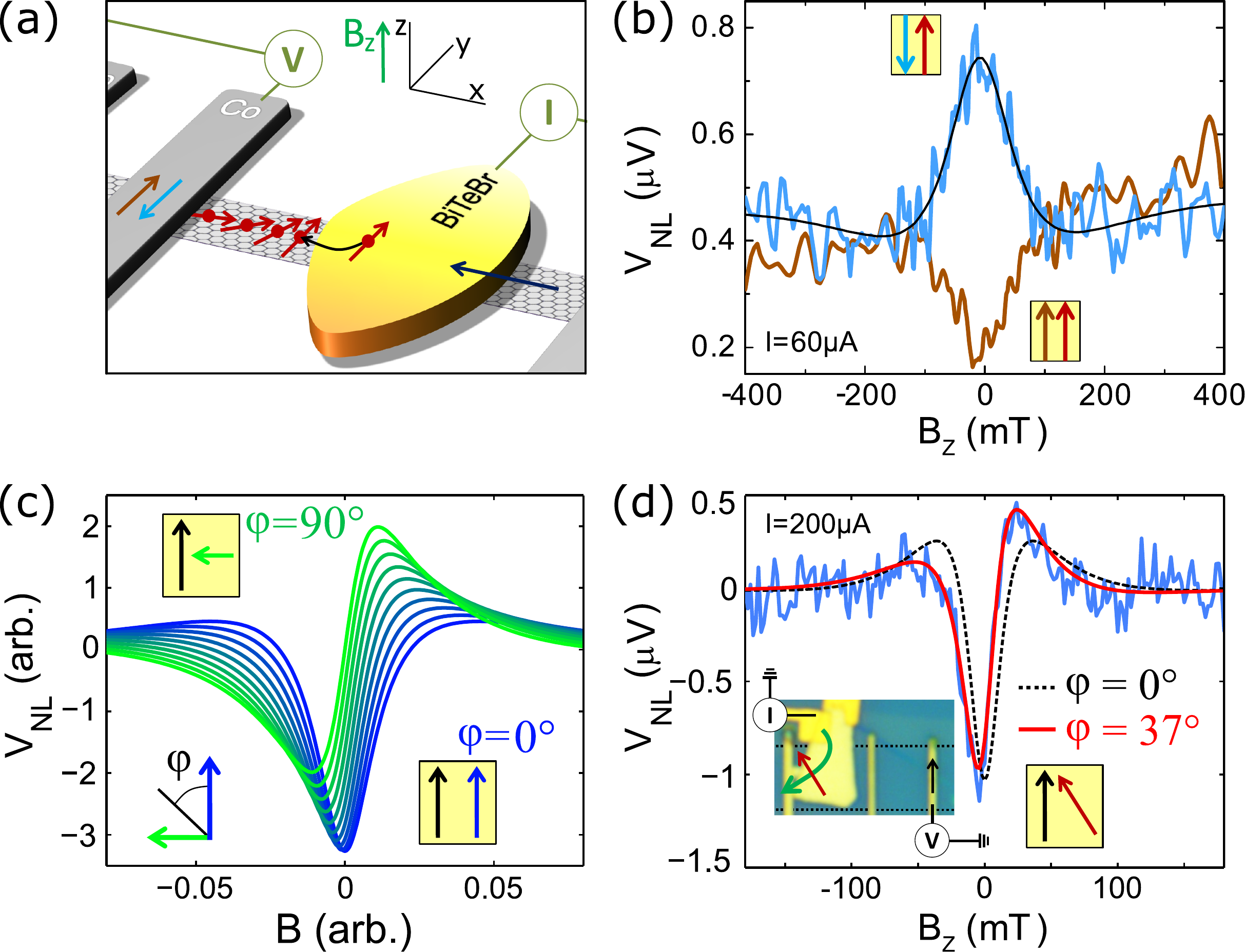}
    \caption{\textbf{Hanle spin precession of injected spin polarization from BiTeBr.} (a) Schematic of device with spin injection from BiTeBr in out-of-plane magnetic field $B_\text{Z}$ resulting in Hanle spin precession in the graphene. (b) Hanle spin precession curves as measured using the setup in (a) with a 60 $\upmu$A bias current, showing both parallel (brown) and antiparallel (blue) configurations of the FM contact and injected spin. There is a slight asymmetry in the signal, indicating an approximately 6{\degree} injector-detector in-plane offset angle. (c) Theoretical Hanle curves for different injector-detector offset angles. The symmetric curve (blue) corresponds to the fully parallel configuration, while the antisymmetric curve (green) to the perpendicular configuration, with intermediary curves in increments of 10{\degree}. (d) Hanle measurement on a second device, shown in the inset image. Fitting indicates a greater offset angle of approximately 37{\degree}. For comparison, the expected curve for a 0{\degree} offset is shown as a dashed black line.}
    \label{fig3}
    \end{center}
\end{figure*}

To further support the origin of the signal in \hyperref[fig2]{Fig.~\ref{fig2}} being spin injection from BiTeBr, Hanle spin precession was also performed in this measurement configuration by using an out-of-plane field, $B_\text{Z}$. \hyperref[fig3]{Fig.~\ref{fig3}} (a) is a schematic of the measurement, while (b) shows the Hanle curves obtained for a parallel (brown) and antiparallel (blue) configuration of the FM detector compared to the injected spin. The Hanle curves show similar spin transport characteristics as those previously measured in FM-FM spin valve configuration in graphene (\hyperref[fig1]{Fig.~\ref{fig1}} (e)), indicating that the detected signal is caused by polarized spins which diffuse in the graphene between the BiTeBr crystal and FM detector.

The Hanle curves in \hyperref[fig3]{Fig.~\ref{fig3}} (b) are slightly asymmetric. The asymmetry and offset from $B_\text{Z}=0$ of the maximum of the Hanle curve originates from the finite precession needed to fully align the incoming spins with the detector polarization and achieve maximum signal amplitude. This indicates an in-plane offset angle between the injected spin and FM detector. \hyperref[fig3]{Fig.~\ref{fig3}} (c) shows the general effect of an injector-detector offset angle on the Hanle spin precession (see Supp. Info.), with the curve smoothly shifting from a symmetric (blue) to an antisymmetric (green) one, in increments of 10 degrees. By fitting, a small angle of 6{\degree} is obtained for the data in \hyperref[fig3]{Fig.~\ref{fig3}} (b). We also present a measurement in another device (Device 2), in \hyperref[fig3]{Fig.~\ref{fig3}} (d), where a more noticeable offset of approximately 37{\degree} is obtained. The device is shown in the inset, where the BiTeBr crystal features Au top contacts. Dotted lines represent the extent of graphene in the device. On the dataset, the red curve is the asymmetric fit, while a reference curve of 0{\degree} offset is shown in dashed black, to visually emphasize the difference.

In case of an FM contact, the orientation of injected or detected spin depends on the magnetization, typically along the easy axis along the length of the FM contact. On the other hand, when injecting using the BiTeBr crystal due to REE, polarization will be perpendicular to the electric field driving charge transport. The BiTeBr crystals have relatively small thickness (100 nm) compared to lateral size (few $\upmu$m), as well as a low resistance compared to the graphene and interface resistance. This suggests that the electric field within them will be predominantly in-plane. This assumption is also supported by finite element simulation on a simplified geometric model of our devices (see Supp. Info.). However, the orientation of the electric field within the xy-plane will depend on the geometry of the irregularly shaped crystal, the position of the metallic top contact and that of the BiTeBr-graphene interface. For Device 2, on the inset in \hyperref[fig3]{Fig.~\ref{fig3}} (d), the green curved arrow depicts the expected current flow and electric field lines in BiTeBr, determining the angle of injected spins (shown in red).

\begin{figure*}[!ht]
    \begin{center}
    \includegraphics[width=1.6\columnwidth]{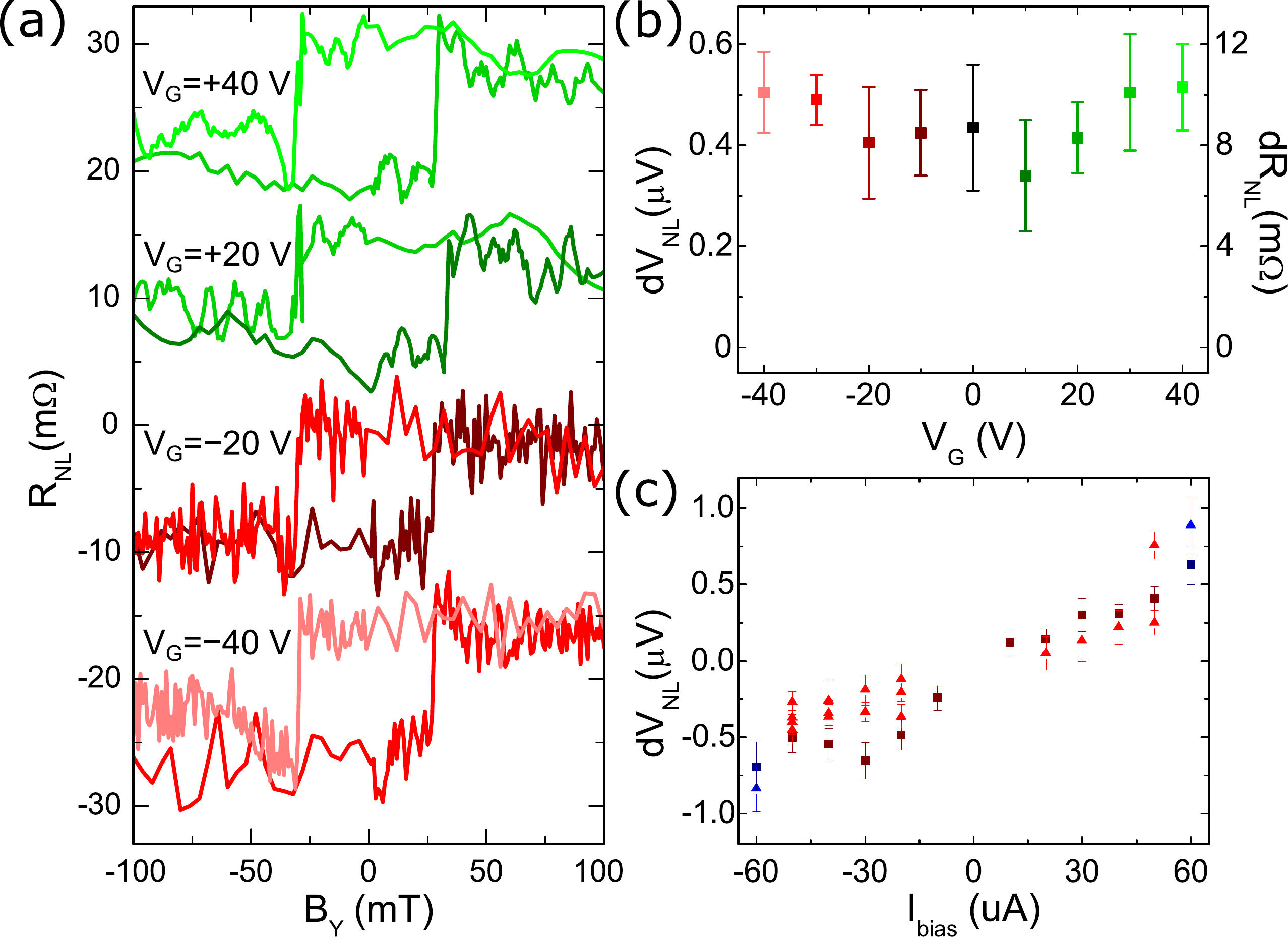}
    \caption{\textbf{Gate and bias dependence of injected spin signal from BiTeBr.} (a) Non-local spin signal switching using an in-plane field $B_\text{Y}$, at various backgate voltages, showing modest change in signal amplitude $dR_\text{NL}$. The data traces are manually stacked using 30\,m$\Upomega$ offsets for better visibility. (b) The $dR_\text{NL}$ values plotted as a function of backgate voltage. The signal has the same orientation under both electron and hole dominated transport in graphene. (c) DC bias current dependence of the spin injection signal amplitude using FM detector contacts D (blue) and F (red) of Device 1.  Triangles represent information extracted from Hanle spin precession, and rectangles represent non-local spin signal data for the same respective sections.}
    \label{fig4}
    \end{center}
\end{figure*}

To further characterize the spin signal injected from BiTeBr, \hyperref[fig4]{Fig.~\ref{fig4}} (a) shows non-local spin valve switching of the signal, measured at different values of the backgate voltage, ranging from -40 to 40\,V, where the curves are offset in y-axis for clarity. Since the charge neutrality point of the graphene section is at approximately 6\,V, this demonstrates that the observed transition does not change sign while transport changes from electron-like carriers to hole-like ones, and cannot be attributed to local Hall effect in FM detector contact. The signal amplitude is seen to change very little with backgate voltage, with values from additional measurements represented in panel (b). This is not surprising, considering the BiTeBr has relatively high charge carrier density, preventing significant gate dependence of resistance, and the resistance of the BiTeBr-graphene interface is also observed to change by no more than a factor of two over this gate voltage range. \hyperref[fig4]{Fig.~\ref{fig4}} (c) depicts a bias current dependence of the amplitude of signals observed on both contacts D (blue) and F (red). Signal amplitude is extracted from both NL spin valve measurements (rectangles) as well as out-of-plane Hanle spin precession measurements (triangles). For the latter, values equal to double the Hanle peak amplitudes were plotted, to correspond with the NL spin valve amplitudes. Note the sign change in the signal with change of bias direction, as also seen in \hyperref[fig2]{Fig.~\ref{fig2}}. This can be explained by REE-induced non-equilibrium between spin-dependent chemical potentials in BiTeBr. Further discussion can be found in the Supp. Info..

The lack of SOI enhancement in graphene under the BiTeBr crystal, the single switching observed in NL measurements while injecting from BiTeBr, the observation of Hanle spin precession and the gate dependence of the NL signal all support the notion that the spin polarization detected in graphene originates from the bulk of BiTeBr. We have also attempted to detect a NL signal in graphene while passing bias current through the BiTeBr crystal only (using two metallic top contacts), but we could not detect a similar switching signal in this case. This is also consistent with the large BiTeBr/graphene interface preventing diffusion of spins from one material to the other.

We can treat the BiTeBr-interface-graphene structure similarly to a FM-tunnel barrier-graphene structure and calculate an interfacial spin polarization\,\cite{pcalc-valet-fert}. Using bias currents of between 40 - 60 $\upmu$A, the experimentally observed polarization in Device 1 is 0.09\,\% and 0.07\,\% when detecting on contact D and F, respectively, and 0.08\,\% in Device 2 while using a bias current of 80 $\upmu$A. We have constructed a tunneling model (detailed in Supp. Info.) to calculate the expected current polarization of electrons - having net spin polarization due to REE - tunneling from BiTeBr into graphene. The model takes into account the 3D band structure of BiTeBr, the position of the Fermi level, $\mu_\text{BiTeBr}\approx50$\,meV, estimated from transport data, and the shift in occupation of electron states in BiTeBr due to the internal electric field, $E_\text{IP}$. For the Rashba parameter, $\alpha_\text{R}\approx 2$\,eV\AA\, was used\,\cite{bitebr-arpes}. The momentum relaxation time $\tau$ can be approximated from the Drude model to be $5.3\times10^{-14}\,$s, reasonably close to that obtained in Ref.\,\cite{bitebr-effmass-n3d-ideue}. $E_\text{IP}$ can be estimated from BiTeBr resistivity and shape, and the bias current. For Device 1, $E_\text{IP}$ fields of approximately 3500 V/m and 3000 V/m are obtained when detecting on contact D and F respectively. Using these values, the tunneling model predicts a spin polarization of the injected current of $P=0.095$\,\% and 0.083\,\% for detecting on contact D and F respectively, which is in good agreement with the experimentally observed values. We see the same agreement for the results obtained in Device 2 as well.

Aside from REE, a competing phenomenon that could result in spin injection into graphene is the Spin Hall Effect (SHE)\,\cite{she-th-vignale,she-th-sinova} also taking place in bulk BiTeBr, caused by the strong intrinsic SOI. The SHE has been studied extensively in TMD/graphene heterostructures\,\cite{gr-tmdc-she-ws2,gr-tmdc-she-theory,gr-tmdc-she-mos2,gr-tmdc-she-wte2,milletari-she-theory-2017}. In case of SHE, the same electric field $E_\text{IP}$ is expected to create spin currents along the z-direction, resulting in spin accumulation at the top and bottom of the BiTeBr crystal. The orientation of spins would also be perpendicular to $E_\text{IP}$, as is the case with REE. We estimate the expected SHE polarization following Ref.\,\cite{dyakonov2010-she} (detailed in Supp. Info.). Using the experimentally observed polarization values of $\sim$\,0.1\,\%, and taking into consideration the measurement uncertainties, we obtain an estimated range for the BiTeBr spin Hall angle, $\alpha_\text{SHE}$, to be within 1.11-5.71. However, because SHE describes a conversion from charge current to spin current, $\alpha_\text{SHE}$ should be limited to $\left |\alpha_\text{SHE}  \right | \leq 1$. Since even the lower bound of our estimate range is over-unity, we consider that REE is a more likely explanation for our experimental results.

The REE and its inverse has been experimentally explored in interfaces between thin films where z-symmetry is broken\,\cite{ree-exp-biag,ree-exp-cubi,ree-exp-cuagbi2o3,ree-exp-remr,du-ree-ptco-2018}, and also more recently in graphene heterostructures. In these cases, the REE mechanism originates from \textit{proximity} induced SOI in graphene at the interface with another material, such as WS$_2$\,\cite{ree-exp-ws2-bart}, WSe$_2$\,\cite{avsar-wse2-opto-ree}, MoS$_2$\,\cite{gr-tmdc-she-mos2}, MoTe$_2$\,\cite{ree-exp-mote2-hoque,safeer-ree-mote-2019}, TaS$_2$\,\cite{Li_tas2_2019} or topological insulators\,\cite{Rodriguez_Vega-ree-2017,khokhri-ree-2019}. In contrast, in our work, due to a large interface resistance, there is no proximity SOI enhancement in the graphene, and the observed spin polarization originates from REE within the BiTeBr \textit{bulk}. The lack of a sign change in the gate dependence of the signal further supports this, as proximity induced REE in graphene is expected to give rise to a sign change as the carrier density is tuned from the electron to the hole region\,\cite{Dyrda-ree-gr-sign-2014,khokhri-ree-2019,ree-exp-mote2-hoque,offidani-signchange-2017}. In addition, we did not observe any sign of weak antilocalization in our devices at temperatures down to 50 mK.

In our tunneling model, the REE induced current polarization  is inversely proportional to the Fermi level in the BiTeBr band, which is in agreement with REE theory\,\cite{ree-edelstein}.  Thus the  polarization should strongly depend on the 3D electron density in BiTeBr, i.e.  $P\sim n^{-2/3}$. The used BiTeBr crystals feature a very high charge carrier density, with a Fermi energy of around 50 meV. Further development of bulk crystal growth techniques could result in lower carrier densities, which leads to an increase in polarization magnitude. Existing works already show a variation of approximately one order of magnitude in the carrier density of bulk crystals\,\cite{bitebr-effmass-n3d-ideue,bitei-ndensity-kanou}. An improvement of one order of magnitude would already produce polarizations comparable to our FM contacts, thus allowing FM injectors to be replaced by BiTeBr. Another promising alternative could be provided by few-layer BiTeBr, allowing for effective gating. Recently a method for exfoliating single-layer BiTeI flakes has been reported\,\cite{sl-bitei}.

\section{Conclusion}

We have demonstrated the electrical creation and control of spin polarization in the giant Rashba spin-orbit crystal BiTeBr at room temperature. Application of an electric bias generates spin polarization in the bulk bands of BiTeBr due to the Rashba-Edelstein effect, where the magnitude and direction are determined by the electric current strength and direction. This spin polarization in BiTeBr is demonstrated by injecting into the graphene channel and detecting in a spin valve device utilizing reliable non-local spin transport and Hanle spin precession measurements. The detailed measurement of the spin signal with different bias current directions and gate voltages proves the robustness of the spin polarization, which is in agreement with current-induced spin polarization from the bulk Rashba spin-split bands of BiTeBr crystal. These findings prove that Rashba spin-orbit crystals are an attractive novel building block for various spintronic applications since they can serve as an all-electrically controlled spin polarization source. Further enhancement and tuning of the current-induced spin polarization is within reach by controlling the Fermi-level position with doping. These advances in electrical control and tunability of spin sources will open new avenues to replace ferromagnetic components in integrated spintronic memory and logic technologies.

\section*{Author Contributions}
Z.K.K. and B.S. fabricated the devices, with help from A.M.H, B.F. and M.K.. The growth and primary characterization of the BiTeBr single crystals was carried out by K.A.K., O.E.T., T.V.K. and M.V.Y.. The hBN crystals were provided by T.T. and K.W.. Z.K.K., B.S. and A.M.H. performed the measurements. Data analysis was performed by Z.K.K., B.S., A.M.H., P.M. and S.C.. Tunneling model was created by B.S. with help from P.M., S.C. and M.K..  All authors contributed to the manuscript and discussions. S.C., S.P.D. and P.M. planned and guided the project.

\section*{Acknowledgments}
Authors thank D. Khokhiriakov and B. Karpiak for their help in device fabrication and measurements, M. G. Beckerné, F. Fülöp, M. Hajdu for their technical support, and T. Fehér, L. Oroszlány, C. Schönenberger, S. O. Valenzuela, A. Virosztek and I. Zutic for useful discussions.
This work has received funding and support from Topograph, CA16218 by COST, the Flag-ERA iSpinText project, the ÚNKP-19-3-II-BME-303 New National Excellence Program of the Ministry of Human Capacities, from the OTKA FK-123894 and OTKA NN-127900 grants, and RFBR project number 19-29-12061. P.M. acknowledges support from the Bolyai Fellowship, the Marie Curie grant and the National Research, Development and Innovation Fund of Hungary within the Quantum Technology National Excellence Program (Project Nr. 2017-1.2.1-NKP-2017-00001). S.P.D. acknowledges funding from Swedish Research Council VR No. 2015-06813 and 2016-03658. M.V.Y. and T.V.K. were supported by the Ministry of Science and Higher Education of the Russian Federation ('Spin' No AAAA-A18-118020290104-2) whereas O.E.T. and K.A.K. were supported by the Russian Science Foundation (No 17-12-01047) and Saint Petersburg State University (Project ID 51126254). K.W. and T.T. acknowledge support from the Elemental Strategy Initiative conducted by the MEXT, Japan and the CREST (JPMJCR15F3), JST.

\bibliography{bibl_Electrical-control-injection-BiTeBr_KovacsKrausz_et_al}
\end{document}